\documentclass[twocolumn,showpacs,prb]{revtex4}%
\usepackage{graphicx}
\begin{document}
\author{O. Leenaerts}
\email{ortwin.leenaerts@ua.ac.be} \affiliation{Universiteit Antwerpen, Departement Fysica,
Groenenborgerlaan 171, B-2020 Antwerpen, Belgium}
\author{B. Partoens}
\email{bart.partoens@ua.ac.be} \affiliation{Universiteit Antwerpen, Departement Fysica,
Groenenborgerlaan 171, B-2020 Antwerpen, Belgium}
\author{F. M. Peeters}
\email{francois.peeters@ua.ac.be} \affiliation{Universiteit Antwerpen, Departement Fysica,
Groenenborgerlaan 171, B-2020 Antwerpen, Belgium}
\date{\today}
\title{Adsorption of H$_2$O, NH$_3$, CO, NO$_2$, and NO on graphene: A first-principles study}

\begin{abstract}
Motivated by the recent realization of graphene sensors to detect individual gas molecules, we investigate the adsorption of H$_2$O, NH$_3$, CO, NO$_2$, and NO on a graphene substrate using
first principles calculations. The optimal adsorption position and orientation of these molecules on
the graphene surface is determined and the adsorption energies are calculated. Molecular doping, i.e. charge transfer between the molecules and the graphene surface, is discussed in light of the density of states and the molecular orbitals of the adsorbates.
The efficiency of doping of the different molecules is determined and the influence of their magnetic moment is discussed.
\end{abstract}

 \maketitle

\section{Introduction}
The synthesis of monolayer graphite (i.e. graphene)~\cite{geim} and the experimental observation of Dirac
charge carriers in this system~\cite{novoselov,zhang} have awakened an enormous interest in this two-dimensional material. The unusual properties of carriers in graphene are a consequence of the gapless and
approximately linear electron dispersion at the vicinity of the Fermi level at two inequivalent
points of the Brillouin zone. In the low-energy limit the quasiparticles in these systems are
described in terms of massless chiral relativistic fermions governed by the Dirac equation.

The good sensor properties of carbon nanotubes are already known for some time,\cite{kong} but recently, the possibility to use graphene as a highly sensitive gas sensor was also reported.\cite{schedin} It was
shown that the increase in graphene charge carrier concentration induced by adsorbed gas molecules can be used
to make highly-sensitive sensors, even with the possibility of detecting individual
molecules. The sensor property is based on changes in the resistivity due to
molecules adsorbed on the graphene sheet that act as donors or acceptors. The sensitivity of
NH$_3$, CO, and H$_2$O up to 1 part per billion was demonstrated, and even the ultimate sensitivity
of an individual molecule was suggested for NO$_2$. These excellent sensor properties of graphene are
due to two important facts: i) graphene is a two dimensional crystal with only a surface and no
volume, which maximizes the effect of surface dopants, and ii) graphene is highly conductive
and shows metallic conductance even in the limit of zero carrier density.

To fully exploit the possibilities of graphene sensors, it is important to understand the
interaction between the graphene surface and the adsorbate molecules. We perform in this
letter first principles calculations for the molecules NH$_3$, NO$_2$, NO, CO, and H$_2$O adsorbed on
graphene. We determine their exact orientation on the surface and their preferential
binding site by calculating their binding energy. Their charge transfer to the graphene surface is
investigated in order to determine the donor or acceptor character of the molecular dopant. 

\section{Computational details}
\begin{figure}[h]
  \centering
\includegraphics[width= 2.5 in]{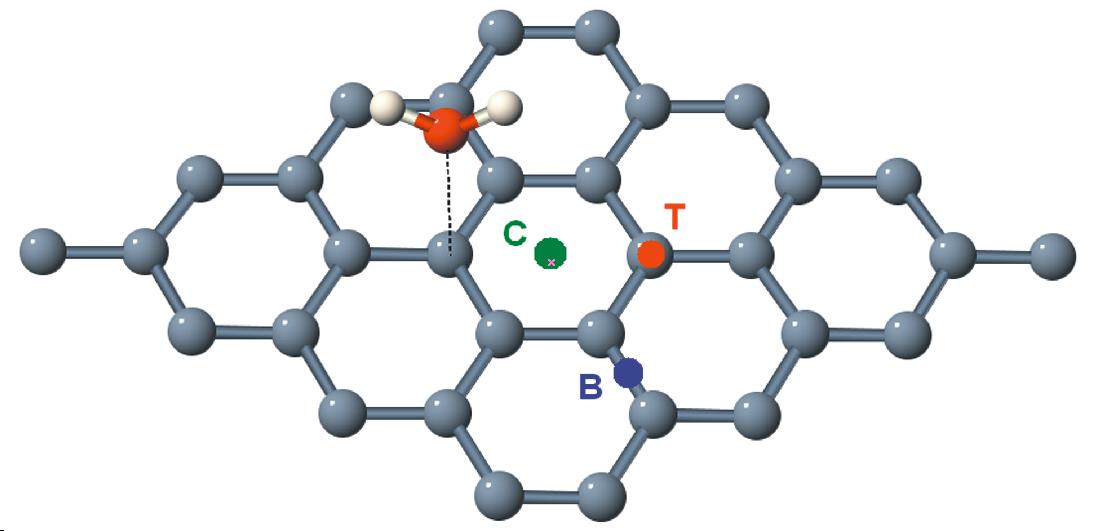}
\caption{\label{fig_h2ogr}(Color online) H$_2$O on graphene. $4\times4$ supercell of graphene with adsorbed H$_2$O molecule.}
\end{figure}

The first principles calculations are performed using density funcional theory (DFT) which has been succesfully used for the study of molecular adsorbates on single-walled (carbon) nanotubes (SWNT).\cite{zhao,santucci,robinson,peng,peng2}
All our DFT calculations were carried out with the Abinit code,\cite{abinitwebsite} within the generalized gradient approximation (GGA) of Perdew, Burke, and Ernzerhof (PBE).\cite{perdew} The advantage of GGA over the local density approximation (LDA) in this work is that the GGA will not lead to a strong bonding of the molecules as in LDA. So if the molecules bind in GGA, they will definitely bind in a real system (and in LDA) too. The distance between adsorbate and the graphene surface, however, will be somewhat overestimated and consequently the binding energy will be underestimated.\\ We use a plane wave basis set with a cutoff energy of 816 eV and pseudopotentials of the Troullier-Martins type.\cite{troullier} For the adsorption of the molecules NH$_3$, CO, and H$_2$O we use non-spin-polarized calculations, while for NO$_2$ and NO, we use spin-polarized ones. The total system consists of a $4\times 4$ graphene supercell (32 C atoms) with a single molecule adsorbed to it (Fig.\ \ref{fig_h2ogr}) and with a distance of 16 {\AA} between adjacant graphene layers. The sampling of the Brillouin zone is done using a $5\times5\times1$ Monkhorst-Pack\cite{monkhorst} grid. For the calculation of the density of states (DOS) we use a $15\times15\times1$ Monkhorst-Pack grid and a Gaussian smearing of 0.14 eV.

Charge transfers are calculated based on the Hirshfeld charge
analysis.\cite{hirshfeld} The atomic charge $Q_A$ for each atom is obtained by (with $\rho({\mathbf r})$ the
calculated density and $\rho_A^0({\mathbf r})$ the electron density computed for the isolated atom
$A$ and taken from Ref.\ \onlinecite{abinitwebsite})
\[
Q_A =\int{ \frac{\rho_A^0({\mathbf r})}{\sum_{A'} \rho_{A'}^0({\mathbf r}) }
\rho({\mathbf r})d\mathbf r},
\]
from which the charge transfer ($\Delta Q$) is deduced. From this result we determined whether or not the adsorbate
 acts as an acceptor or a donor. It should be noted that the size of the charge transfer is slightly dependent on the method used to calculate it.\\
The distance from the adsorbate to the graphene surface is calculated from the difference in weighted averages of the different atoms of the molecule and the carbon atoms of the graphene sheet, where we used the atomic number Z of the atoms as the weight factor.

\section{Results}
For each adsorbate three adsorption sites are considered, namely on top of a carbon atom (T), the center of a carbon hexagon (C) and  the center of a carbon-carbon bond (B) (see Fig. 1) . For these positions, different orientations of the molecules are examined and the adsorption energy is calculated for all of them. The adsorption energy (E$_a$) is the energy of the isolated
graphene sheet and isolated molecule minus the energy of the fully relaxed graphene sheet with the
molecule adsorbed to it.  The strength of the molecular doping is discussed in light of the density of states and  the highest occupied  and lowest unoccupied molecular orbitals (HOMO and LUMO) of the adsorbate. The position of these orbitals, visible as peaks in the DOS, is practically independent of the orientation and adsorption site of the molecule, so we only show the total DOS for one geometry per molecule. We can now distinguish two charge transfer mechanisms: i) a charge transfer can occur due to the relative position in the DOS of the HOMO and LUMO of the adsorbate. If the HOMO is above the Fermi level of pure graphene (the Dirac point), there is a charge transfer to graphene. If  the LUMO is  below the Dirac point, charge will transfer to the molecule. ii) The charge transfer between adsorbate and graphene is also partially determined by the mixing of the HOMO and LUMO  with the graphene orbitals (hybridisation). This mixing scales with the overlap of the interacting orbitals and the inverse of their energy difference.   \\
It is more difficult to discuss the adsorption energy in this way because of the large amount of possible interacting orbitals present in graphene.
Our investigation starts with the non-magnetic molecules H$_2$O, NH$_3$, and CO, followed by the paramagnetic ones, NO$_2$ and NO. We discuss the molecules in the order of increasing complexity of their charge transfer mechanism.

\subsection{H$_2$O on graphene}
We examine the following  orientations of the H$_2$O molecule with respect to the graphene surface: starting from the O atom the H-O bonds pointing up (u), down (d) or parallel to the graphene surface (n). Another orientation (v) was suggested in a theoretical study, based on an empirical method, of the adsorption of H$_2$O on graphite.\cite{gonzalez}  This orientation has one O-H bond parallel to the surface and the other one pointing to the surface. All properties were found to be almost invariant with respect to rotations around the axis perpendicular to the surface and through the oxygen atom, and therefore we will not discuss this orientation. The results of the calculations are given in table~\ref{tab-h2o}.

\begin{table}[h]
\caption{H$_2$O on graphene: the adsorption energy ($E_a$), the distance of H$_2$O above the graphene surface ($d$), and the charge transfer from the molecule to graphene ($\Delta Q$) for ten different geometries.\label{tab-h2o}}
\begin{tabular}{ccccc}
\hline\hline
{Position} & { Orientation} & { $E_a$(meV)} & \hspace{1mm} { $d$(\AA)} \hspace{1mm} &  {$\Delta Q$(e)} \\
\hline
         B &          u &       18.4 &       3.70 &     0.021 \\

         T &          u &       18.7 &       3.70 &     0.021 \\

         C &          u &       20.3 &       3.69 &     0.021 \\

         B &          n &       23.8 &       3.55 &     0.013 \\

         T &          n &       23.7 &       3.56 &     0.015 \\

         C &          n &       26.5 &       3.55 &     0.014 \\

         B &          d &       17.8 &       4.05 &    -0.009 \\

         T &          d &       18.5 &       4.05 &    -0.009 \\

         C &          d &       19.4 &       4.02 &    -0.010 \\

    {\bf C} &    {\bf v} & {\bf 47.0} & {\bf 3.50} & {\bf -0.025} \\

\hline\hline
\end{tabular}

\end{table}

\begin{figure}[h!]
  \centering
\includegraphics[width= 3.375 in]{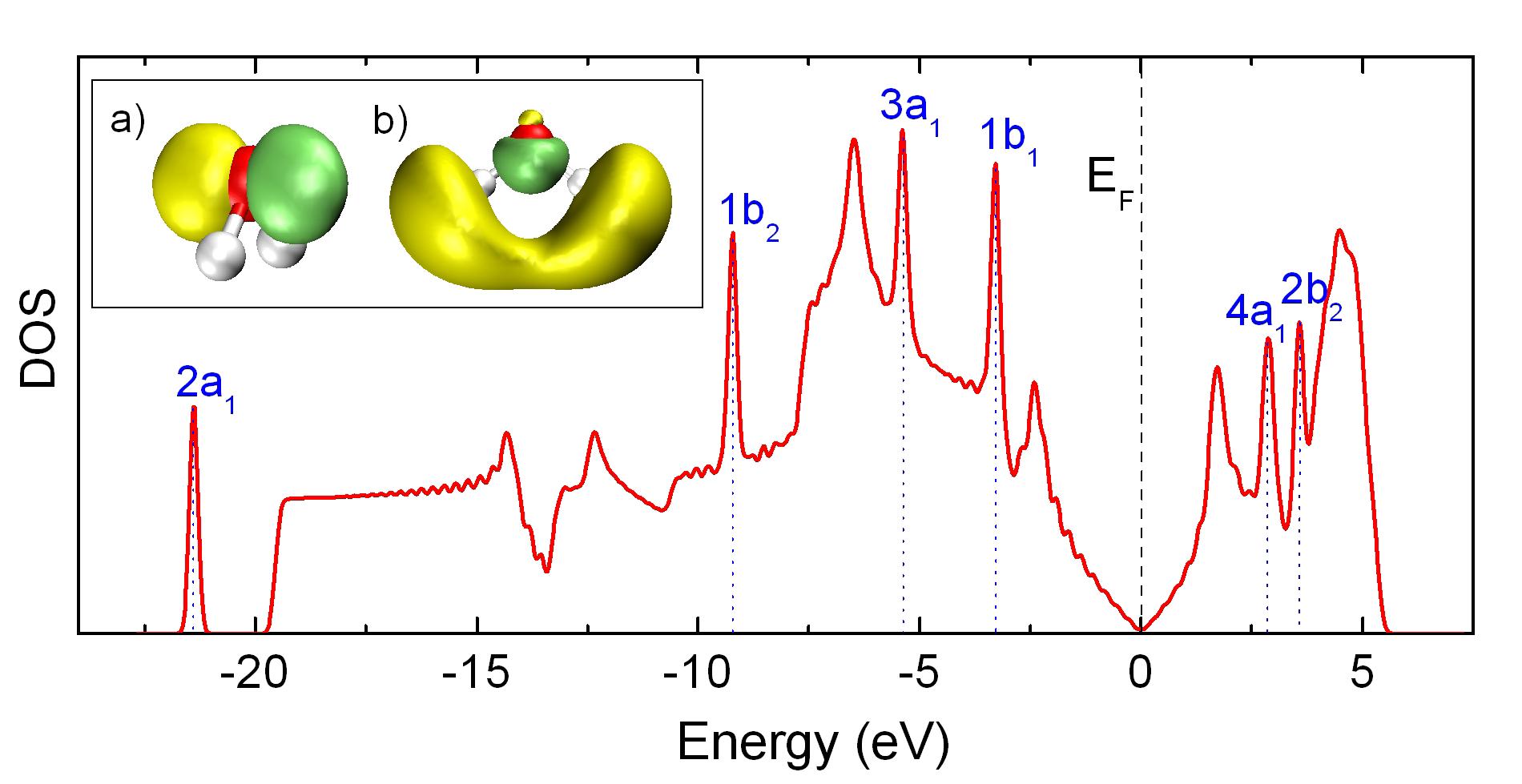}
\caption{(Color online) H$_2$O on graphene. Inset: (a) the HOMO and (b) the LUMO of H$_2$O (the H atoms are white and the oxygen atom is red; green and yellow indicate different signs of the orbital wavefunction). Main panel: DOS of H$_2$O on graphene. The blue dotted lines show the positions of the molecular orbitals of H$_2$O.\label{fig:h2ogr}}
\end{figure}

From table~\ref{tab-h2o} we learn that all the adsorption energies are small, which is partially a consequence of the calculation method. This is not very important because the adsorption energies are only used to compare the different geometries and to find the best position and orientation of the molecule for which we need only relative values. Table 1 also shows that the adsorption energy is primarily determined by the orientation (u, d, n, v) and to a lesser degree by the position (C, B, T) of the molecule. The energy differences are 5-6 meV with respect to the orientation, but they vary with about 1-2 meV when changing the position. This difference in importance of position and orientation is even more pronounced when we look at the charge transfers.  If the O atom points to the graphene surface, there is a (small) charge transfer to graphene, but if the H atoms point to the surface there is a small charge transfer to the water molecule. This is a consequence of the form of the HOMO and LUMO of H$_2$O and their relative position with respect to the Dirac point (see Fig.\ \ref{fig:h2ogr}).

The HOMO (1b$_1$) is completely  located on the O atom, but the LUMO (4a$_1$) is mostly located on the H atoms. In the u orientation the HOMO plays the dominant role and donates, through a small mixing with graphene orbitals above the Fermi level, some charge to graphene. There is also a (stronger) mixing with the orbitals below the Dirac point, because they are closer in energy, but this does not induce any charge transfer because all these orbitals are filled. In the d and v orientation, the LUMO of H$_2$O interacts stronger with the surface and is able, through a small mixing with the graphene orbitals below the Dirac point, to accept some charge from graphene. There is also a stronger mixing with orbitals above the Dirac point now, but this does not induce any charge transfer because all these orbitals are empty. In the n orientation it is again the HOMO that will interact stronger, but now there is also some interaction with the LUMO. There will be a charge transfer from the molecule to graphene, but, because of the interaction with the LUMO, it will be smaller.\\ Experimentally~\cite{schedin} one finds that H$_2$O acts as an acceptor on graphene which is in accordance with our theoretical results where we find  that the acceptor character (C,v) is energetically favored on perfect graphene.

\subsection{NH$_3$ on graphene}

Two orientations of the ammonia molecule were investigated, one with the H atoms pointing away from the surface (u) and the other with the H atoms pointing to the surface (d). All properties were again found to be almost invariant to rotations around the axis perpendicular to the surface and through the nitrogen atom. The results of the calculations are given in table \ref{tab-nh3}. The adsorption site and the orientation are now seen to be of the same importance for the adsorption energy. The charge transfer, however, is solely determined by the orientation of the NH$_3$ molecule.

\begin{table}[h]
\caption{NH$_3$ on graphene: the adsorption energy ($E_a$), the distance of NH$_3$ above the graphene surface ($d$), and the charge transfer from the molecule to graphene ($\Delta Q$) for six different geometries.\label{tab-nh3}}
\begin{tabular}{ccccc}
\hline\hline
{Position} & { Orientation} & { $E_a$(meV)} & \hspace{1mm} { $d$(\AA)} \hspace{1mm} &  {$\Delta Q$(e)} \\
\hline
         B &          u &       21.1 &       3.86 &     0.026 \\

         T &          u &       20.1 &       3.86 &     0.026 \\

   {\bf C} &    {\bf u} & {\bf 30.8} & {\bf 3.81} & {\bf 0.027} \\

         B &          d &       14.7 &       4.08 &     0.001 \\

         T &          d &       15.6 &       3.97 &     0.000 \\

         C &          d &       24.7 &       3.92 &    -0.001 \\
\hline\hline
\end{tabular}

\end{table}

\begin{figure}[h!]
  \centering
\includegraphics[width= 3.375 in]{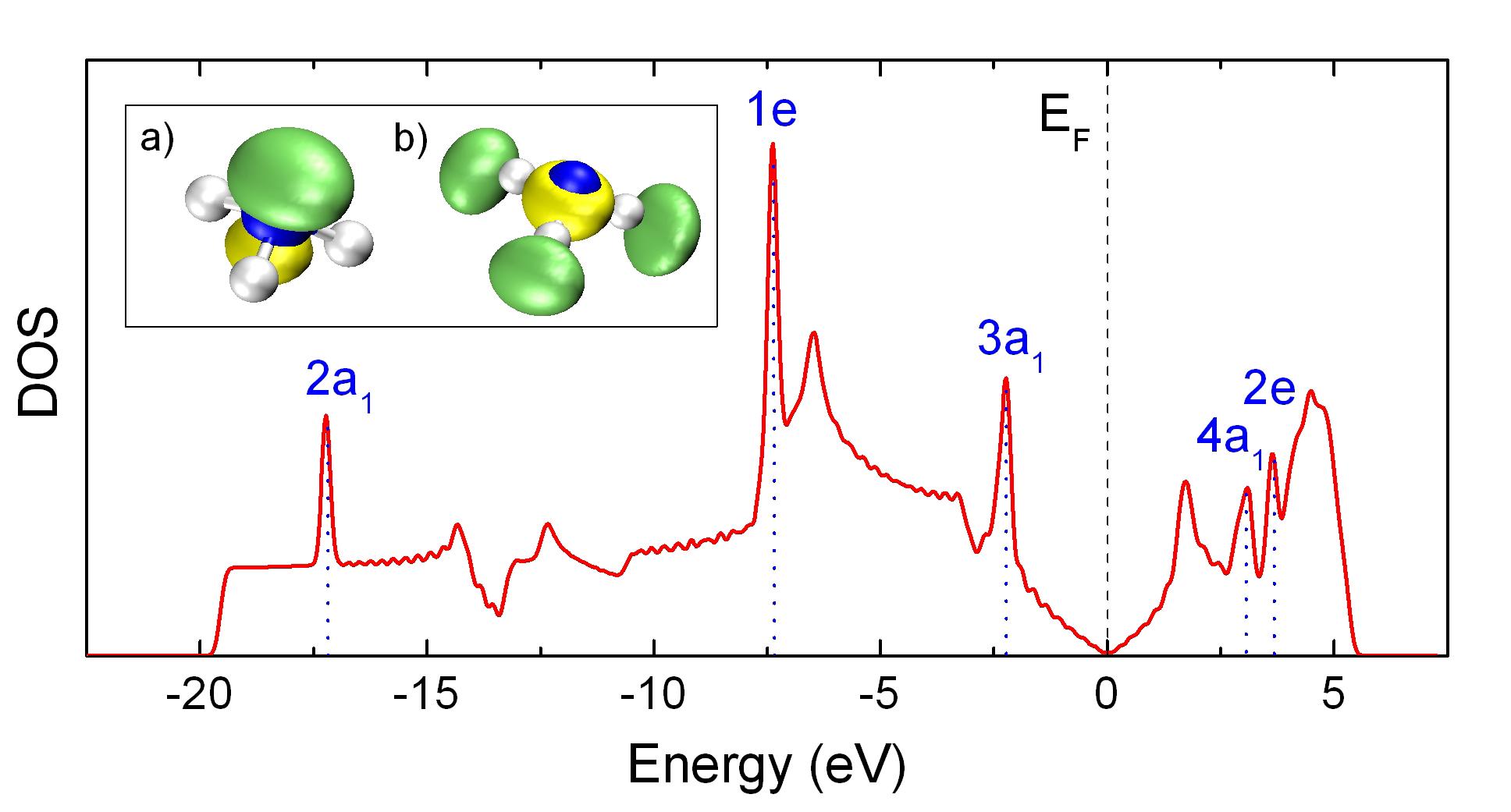}
\caption{(Color online) NH$_3$ on graphene. Inset: (a) the HOMO and (b) the LUMO of NH$_3$ (the N atom is blue and the H atoms are white). Main panel: DOS of NH$_3$ on graphene.\label{fig:nh3gr}}
\end{figure}

There is a small charge transfer from the molecule to the graphene surface of 0.03 e in the u orientation and there is (almost) no charge transfer in the d orientation. We can see how this comes about by looking at the HOMO (3a$_1$) and LUMO (4a$_1$) of the NH$_3$ molecule (Fig.\ \ref{fig:nh3gr}(a) and (b)). In the u orientation, the HOMO is the only orbital that can have a significant overlap with the graphene orbitals and thus can cause charge transfer. As a consequence the NH$_3$ molecule will act as a donor. In the d orientation, both HOMO and LUMO can cause charge transfers which are similar in magnitude but in opposite directions. The net charge transfer is therefore close to 0. \\
The u orientation is energetically favored which explains the donor character as observed experimentally~\cite{schedin}.

We also performed LDA calculations for the adsorption of NH$_3$ on graphene. The results of these are similar for the charge transfer, but the adsorption energy is much larger ($\pm$100 meV). The real adsorption energy is known to lie between the two approximate values obtained through GGA and LDA. This is consistent with the experimental observation~\cite{schedin} that the adsorbates can be removed from the surface by annealing at 150$^\circ$C.

\subsection{CO on graphene}

Three different orientations were used for the CO molecule. Two with the molecule perpendicular to the surface, with the O atom above the C atom (u) and the other way around (d), and one parallel to the surface (n).

\begin{table}[h]
\caption{CO on graphene: the adsorption energy ($E_a$), the distance of CO above the graphene surface ($d$), and the charge transfer from the molecule to graphene ($\Delta Q$) for seven different geometries.\label{tab-co}}
\begin{tabular}{ccccc}
\hline\hline
{Position} & { Orientation} & { $E_a$(meV)} & \hspace{1mm} { $d$(\AA)} \hspace{1mm} &  {$\Delta Q$(e)} \\
\hline
         B &          u &       10.0 &      3.75  &     0.019 \\

         T &          u &        9.6 &      3.75  &     0.019 \\

         C &          u &       13.1 &      3.73  &     0.019 \\

         T &          d &        8.4 &      3.72  &     0.009 \\

         C &          d &        9.6 &      3.70  &     0.010 \\

         B &          n &       14.0 &      3.74  &     0.013 \\

   {\bf C} &    {\bf n} & {\bf 14.1} & {\bf 3.74} & {\bf 0.012} \\
\hline\hline
\end{tabular}

\end{table}

From table \ref{tab-co} we notice that the CO molecule always acts as a donor. The size of the charge transfer only depends on the orientation of the molecule with respect to the surface. The differences in charge transfer are due to differences in orbital overlap between the HOMO ($5\sigma$) of the CO molecule and graphene. The LUMO ($2\pi$) seems to play no important role in the doping process although it is closer to the Dirac point than the HOMO. To understand this we have to take into account the symmetry of this orbital and the graphene orbitals. The DOS below and close (\textless{} 3 eV) to the Dirac point originates from (mostly) bonding combinations of the p$_z$ atomic orbitals of the C atoms of graphene. The DOS above the Dirac point is mostly due to anti-bonding combinations. The completely anti-symmetric LUMO will therefore mostly interact with the DOS above the Dirac point which does not cause any doping.\\
The HOMO is thus the more important orbital and the charge transfer is consequently always to graphene. Because the HOMO is mainly located on the C atom, the charge transfer is the largest when the C atom  is closest to the surface (u orientation), the smallest when the O atom is closer to the surface (d orientation ) and intermediate when both atoms are at an (almost) equal distance from the surface (n orientation).

\begin{figure}[h!]
  \centering
\includegraphics[width= 3.375 in]{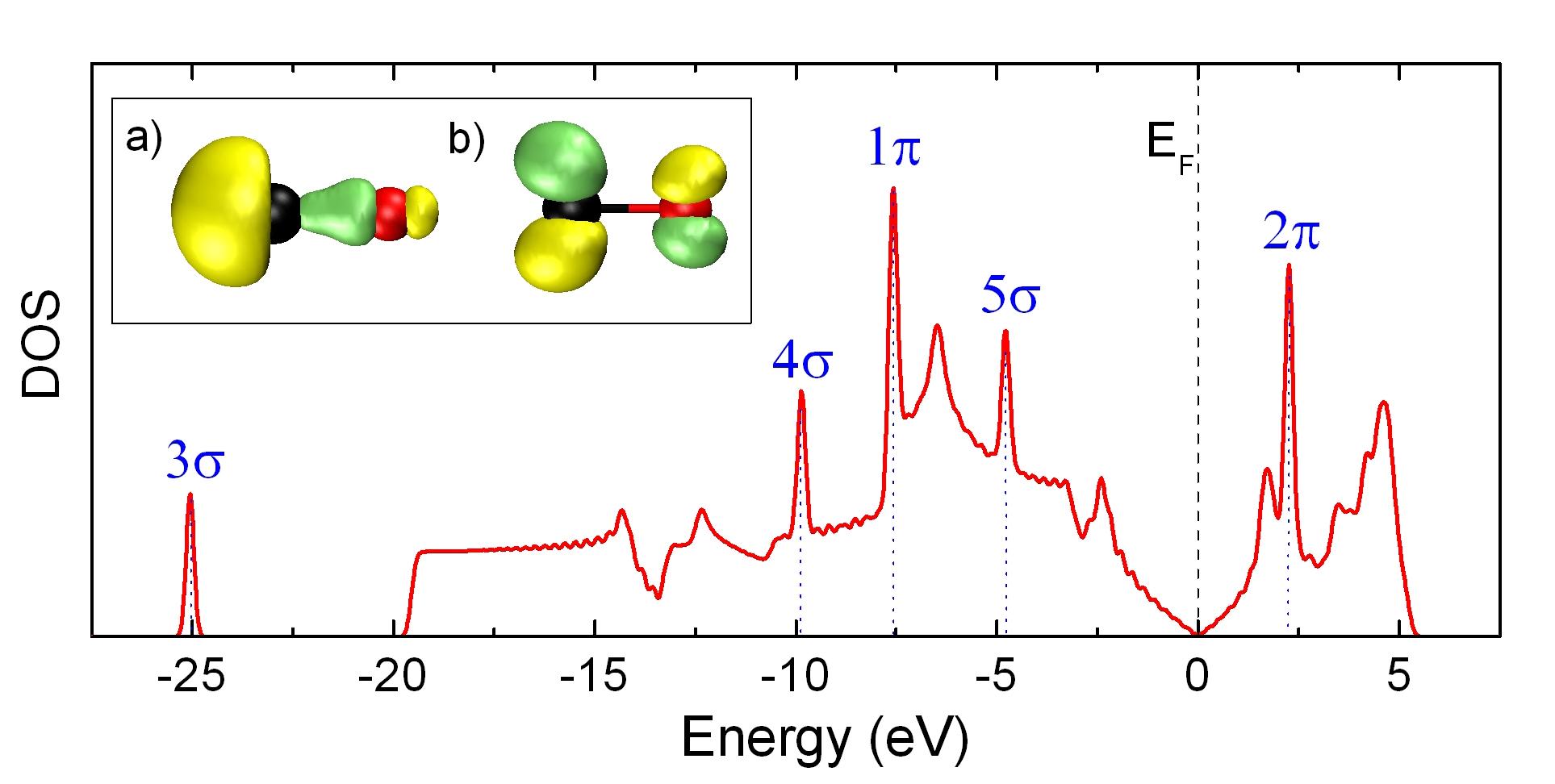}
\caption{(Color online) CO on graphene. Inset: (a) the HOMO and (b) the LUMO of CO (the C atom is black and the O atom is red). Main panel: DOS of CO on graphene.\label{fig:cogr}}
\end{figure}

\subsection{NO$_2$ on graphene}

In Ref.\ \onlinecite{wehling} it was stated that adsorbates with a magnetic moment in general result in a larger doping. Therefore we will now turn our attention to paramagnetic molecules. The first one is NO$_2$ which has, in a spin-polarized calculation, an energy that is 0.4 eV smaller as compared to an unpolarized one and therefore is paramagnetic.
We examine three different orientations of the NO$_2$ molecule: starting from the N atom, the N-O bonds pointing up (u), down (d) or parallel to the graphene surface (n).

\begin{table}[h]
\caption{NO$_2$ on graphene: the adsorption energy ($E_a$), the distance of NO$_2$ above the graphene surface ($d$), and the charge transfer from the molecule to graphene ($\Delta Q$) for six different geometries.\label{tab-no2}}
\begin{tabular}{ccccc}
\hline\hline
{Position} & { Orientation} & { $E_a$(meV)} & \hspace{1mm} { $d$(\AA)} \hspace{1mm} &  {$\Delta Q$(e)} \\
\hline
    {\bf B} &   {\bf d} &  {\bf 67.4} & {\bf 3.61} & {\bf -0.099} \\

         T &          d &       65.3 &       3.61 &      -0.099 \\

         C &          d &       62.6 &       3.64 &      -0.098 \\

         B &          u &       54.7 &       3.83 &      -0.089 \\

         T &          u &       54.5 &       3.93 &      -0.090 \\

         C &          n &       66.7 &       3.83 &      -0.102 \\
\hline\hline
\end{tabular}

\end{table}

\begin{figure}[h]
  \centering
\includegraphics[width= 3.375 in]{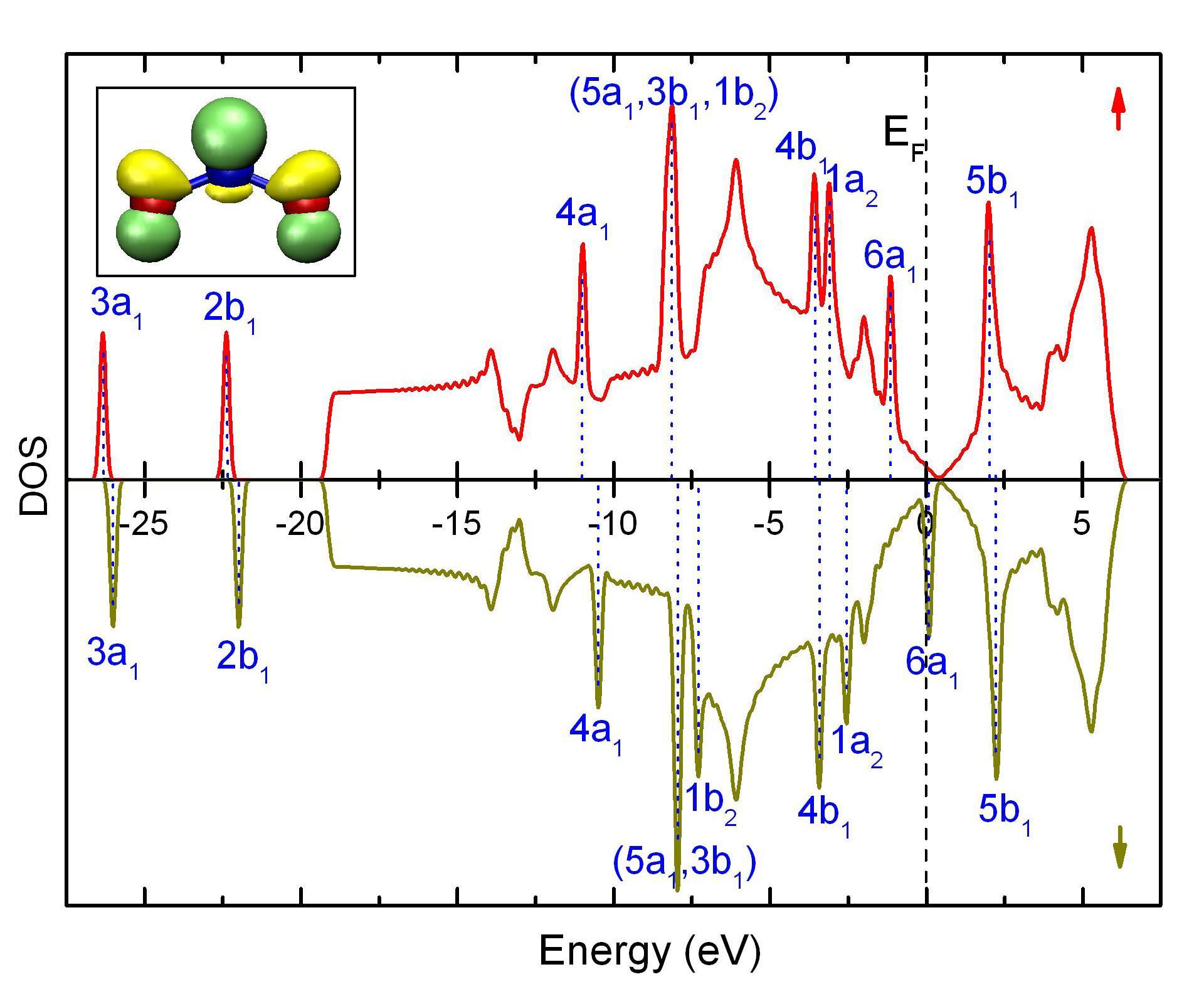}
\caption{(Color online) NO$_2$ on graphene. Inset: HOMO and LUMO of NO$_2$ (the N atom is blue and the O atom is red). Main panel: spin-polarised DOS of NO$_2$ on graphene.\label{fig:no2gr}}
\end{figure}

The LUMO (6a$_1$,$\downarrow$) of NO$_2$ is located 0.3 eV below the Dirac point (Fig.\ \ref{fig:no2gr}). This induces a large charge transfer to the molecule. But there are also some NO$_2$ orbitals close enough to the Dirac point to cause some charge transfer in the opposite direction through orbital mixing (especially the HOMO (6a$_1$,$\uparrow$)). The latter charge transfer is smaller than the first one but it is still noticeable in the strength of the magnetic moment of the system. The total magnetic moment of graphene and adsorbate in e.g. the (B, d) orientation is 0.862 $\mu_b$. The charge transfer from graphene (M=0$\mu_b$) to NO$_2$ (M=1$\mu_b$) is 0.099 e, so the orbital mixing causes a charge transfer of $\pm$0.039 e to graphene.

\subsection{NO on graphene}

\begin{figure}[h]
  \centering
\includegraphics[width= 3.375 in]{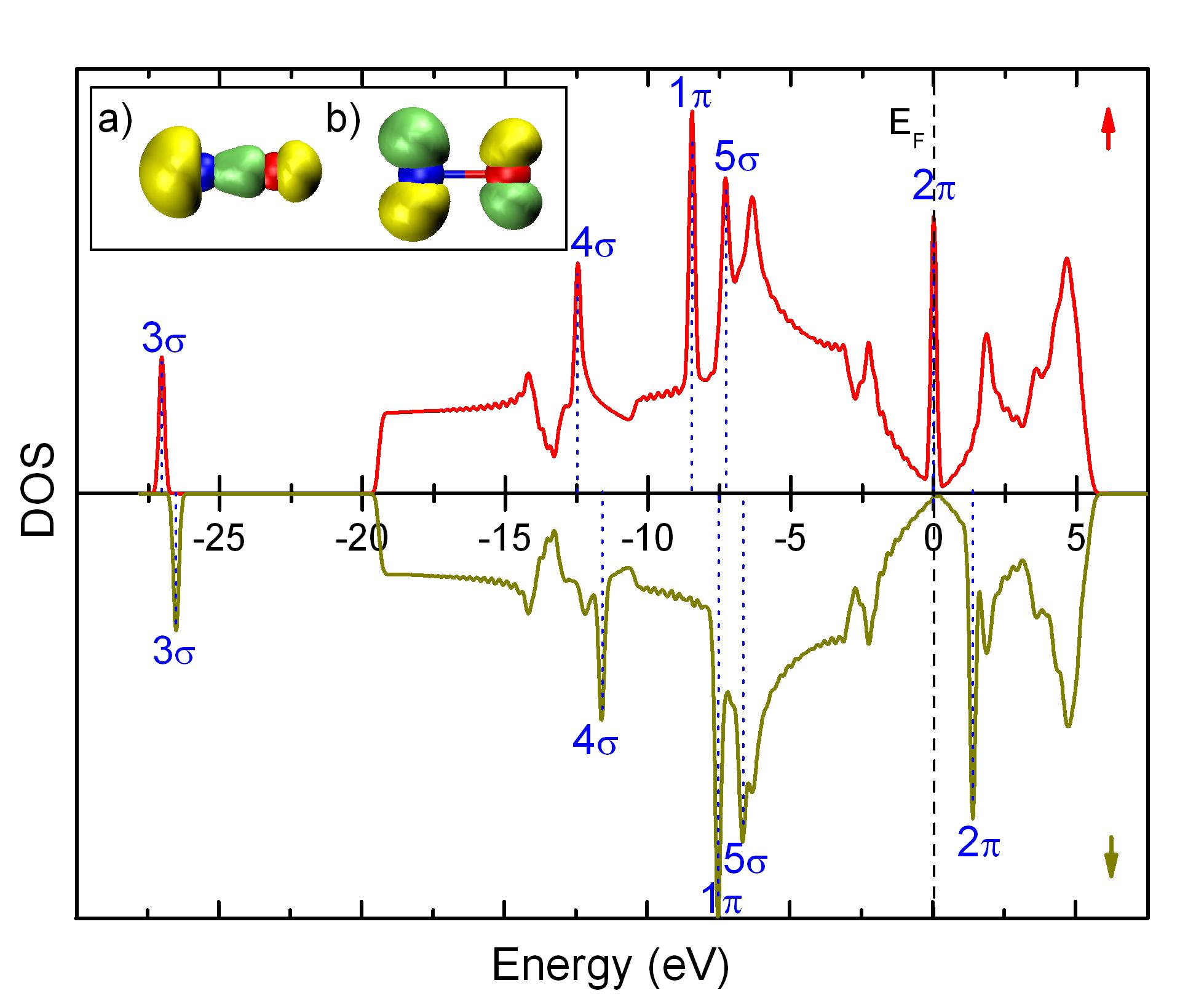}
\caption{(Color online) NO on graphene. Inset: (a) $5\sigma$ orbital and (b) HOMO/LUMO of NO (the N atom is blue and the O atom is red). Main panel: spin-polarized DOS of NO on graphene.\label{fig:nogr}}
\end{figure}

To test whether or not there is always strong doping by paramagnetic
molecules, we will investigate another one. For NO a spin polarized
calculation gives an energy that is 0.3 eV lower than a
non-spin-polarized one, so NO is indeed a paramagnetic molecule. We
investigate the same orientations and use the same notations as for
the CO molecule (replace C with N). Contrary to the claim made in
Ref.\ \onlinecite{wehling}, we did not find that NO induces any
strong doping. The charge transfers are an order of magnitude
smaller than in the case of the NO$_2$ molecule (see table
\ref{tab-no}) which is comparable to the non-magnetic molecules.
Physically, we can understand this if we compare the DOS of the
adsorbates NO (Fig.\ \ref{fig:nogr}) and NO$_2$ on graphene. For
NO$_2$ adsorbed on graphene, the LUMO is situated 0.3 eV below the
Dirac point of graphene. This induces a strong doping. In the case
of NO (Fig.\ \ref{fig:nogr}) the HOMO is degenerate ($2\pi_x$,
$2\pi_y$) and is half filled (so it is also the LUMO) and lies only
0.1 eV below the Dirac point. This induces a very small charge
transfer from graphene to NO, but, due to its small strength, it can
be (over)compensated by orbital mixing. The HOMO/LUMO of NO can,
because it is half filled, cause charge transfer in both directions
by mixing with the graphene orbitals below and above the Dirac
point. But, as in the case of the LUMO of CO, it interacts mostly
with the latter due to symmetry reasons. So the orbital mixing leads
to charge transfer to graphene.

\begin{table}[h]
\caption{NO on graphene: the adsorption energy ($E_a$), the distance of NO above the graphene surface ($d$), and the charge transfer from the molecule to graphene ($\Delta Q$) for six different geometries.\label{tab-no}}
\begin{tabular}{ccccc}
\hline\hline
{Position} & { Orientation} & { $E_a$(meV)} & \hspace{1mm} { $d$(\AA)} \hspace{1mm} &  {$\Delta Q$(e)} \\
\hline
         C &          u &        15.7 &      4.35 &     0.006 \\

         T &          u &        14.0 &      4.35 &     0.006 \\

         C &          d &        12.6 &      4.11 &     0.007 \\

         T &          d &        10.6 &      4.27 &     0.005 \\

         C &          n &        27.9 &      3.71 &     0.018 \\

      {\bf B} & {\bf n} &  {\bf 28.5} & {\bf 3.76} & {\bf 0.017}  \\

\hline\hline
\end{tabular}

\end{table}

We see from table \ref{tab-no} that the charge transfer due to orbital mixing always overcompensates the small transfer due to the position of the HOMO/LUMO, so NO always act as a donor.
We notice that there are large differences in $\Delta Q$ and in the distance d. They obviously correlate because a smaller distance between adsorbate and graphene leads to a larger orbital overlap and consequently to more orbital mixing (i.e. a larger charge transfer).
The differences in the distance can be explained by the overlap of the $5\sigma$ orbital. The position of this orbital is very close in energy to that part of the graphene DOS that originates from the (bonding) combinations of carbon p$_z$ orbitals around the $\Gamma$ point. Mixing of these orbitals induces a net energy shift upwards so they repel each other strongly. The geometry of the $5\sigma$ orbital gives a large overlap in the u orientation, a smaller overlap in the d orientation and the smallest overlap in the n orientation. This  gives a simple explanation for all the differences found from our calculations.

\section{Summary and conclusions}
The charge transfer between the considered adsorbates and graphene is found to be almost independent on the adsorption site but it does depend strongly on the orientation of the adsorbate with respect to the graphene surface. We compared two paramagnetic molecules, NO$_2$ and NO, and found that NO$_2$ induces a relatively strong doping (-0.1e), but NO does not (\textless{}0.02e). This is in contrast to Ref.\ \onlinecite{wehling} where it was claimed that paramagnetic molecules are strong dopants which we found indeed to be the case for NO$_2$ but not so for NO. \\

\begingroup
\squeezetable
\begin{table}[]
\caption{Summary of results.\label{tab-res}}
\begin{tabular}{ccccc}
\hline\hline
{Adsorbate} & { Theory} & { Experiment~\cite{schedin}} & \hspace{1mm} {E$_a$ (meV)} \hspace{1mm}& $\Delta$Q(e) \\
\hline
         H$_2$O &   acceptor &   acceptor &  47 &   -0.025 \\

         NH$_3$ &    donor   &    donor   &  31 &    0.027 \\

         CO     &    donor   &    donor   &  14 &    0.012 \\

         NO$_2$ &   acceptor &   acceptor &  67 &   -0.099 \\

         NO     &    donor   &      /     &  29 &    0.018 \\

\hline\hline
\end{tabular}
\end{table}
\endgroup

For the considered adsorbates the sign of the charge transfer agrees with what was found experimentally (see table \ref{tab-res}) in Ref. \onlinecite{schedin}. Our results are also in good agreement with theoretical studies of the adsorption of molecules on large SWNT's in e.g. Ref. \onlinecite{zhao}. This suggests that some of the knowlegde of adsorption on nanotubes should be transferable to graphene.

\begin{acknowledgments}
This work was supported by the Flemish Science Foundation (FWO-Vl), the NOI-BOF of the University of Antwerp and the Belgian Science Policy (IAP).
\end{acknowledgments}

\end{document}